\documentclass[aps,prb,reprint,floatfix,amsmath,amssymb, superscriptaddress,sort&compress]{revtex4-2}

\usepackage{graphicx}
\usepackage{epstopdf}
\usepackage{color}
\usepackage{hyperref}
\usepackage{bm}
\usepackage{printlen}
\usepackage{units}

\renewcommand{\vec}[1]{\mathbf{#1}}

\begin{document}
	\title{Electronic transport in graphene with out-of-plane disorder}
	\begin{abstract}
Real-world samples of graphene often exhibit various types of out-of-plane disorder--ripples, wrinkles and folds--introduced at the stage of growth and transfer processes. These complex out-of-plane defects resulting from the interplay between self-adhesion of graphene and its bending rigidity inevitably lead to the scattering of charge carriers thus affecting the electronic transport properties of graphene. We address the ballistic charge-carrier transmission across the models of out-of-plane defects using tight-binding and density functional calculations while fully taking into account lattice relaxation effects. The observed transmission oscillations in commensurate graphene wrinkles are attributed to the interference between intra- and interlayer transport channels, while the incommensurate wrinkles show vanishing backscattering and retain the transport properties of flat graphene. The suppression of backscattering reveals the crucial role of lattice commensuration in the electronic transmission. Our results provide guidelines to controlling the transport properties of graphene in presence of this ubiquitous type of disorder.

	\end{abstract}
	
	\author{Yifei Guan}
	\affiliation{Institute of Physics, \'{E}cole Polytechnique F\'{e}d\'{e}rale de Lausanne (EPFL), CH-1015 Lausanne, Switzerland}
	
	\author{Oleg V. Yazyev}
	\email{E-mail: oleg.yazyev@epfl.ch}
	
	\affiliation{Institute of Physics, \'{E}cole Polytechnique F\'{e}d\'{e}rale de Lausanne (EPFL), CH-1015 Lausanne, Switzerland}
	
	\date{\today}
	
	\maketitle
	
\begin{figure}[b]
	\centering
		\includegraphics[width=\linewidth]{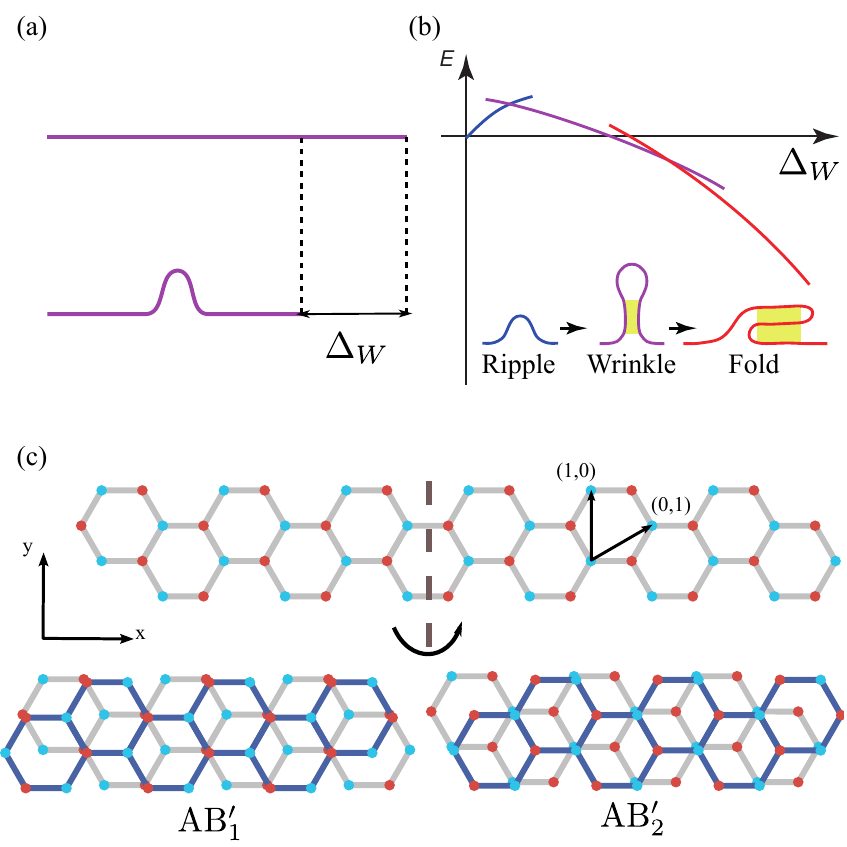}
		
		\caption{The structure of out-of-plane disorder in graphene. 
		(a) Definition of compressive displacement $\Delta_W$ relative to the flat, unstrained graphene.
        (b) Formation of the three distinct types of out-of-plane disorder upon increasing $\Delta_W$. The curves show a schematic illustration of the dependence of energy $E$ on $\Delta_W$ for the three deformation regimes. Yellow color exposes the collapsed regions where the interlayer coupling is enabled.
        (c) Illustration of the interlayer coupling between the atoms belonging to the same sublattice in commensurate zigzag wrinkles and folds.}
		\label{fig:1}
\end{figure}

Being the first and the most investigated two-dimensional (2D) material, graphene continues attracting attention as a platform for exploring novel physics and realizing prospective technological applications~\cite{castro2009electronic}. The 2D nature of graphene gives rise to soft flexural modes that result in low-energy out-of-plane disorder otherwise absent in bulk, three-dimensional materials~\cite{deng2017wrinkle, mariani2008flexural,croy2020bending,de2015soliton}. The interplay between bending upon in-plane compression and the interlayer adhesion results in several distinct types of out-of-plane disorder: ripples, wrinkles and folds (see Refs.~\cite{Zhu2012,deng2017wrinkle} and Figs.~\ref{fig:1}(a,b)). The out-of-plane disorder has a prominent effect on the electronic structure and transport properties of graphene~\cite{hattab2012interplay,xieElectronTransportFolded2012,Pelc2015a,afmat2020wrinklemambr}.
Finite curvature of the deformed region results in pseudo-gauge fields~\cite{vozmediano2008gauge,ortolani2012folded}, while the collapsed regions in wrinkles and folds provide a pathway for electronic tunnelling between layers~\cite{Zhu2012,fernando2015oscillation}. In addition, out-of-plane disorder locally accumulates charges and act as scattering centers~\cite{guo2013electronic,pereira2010geometry,Zhu2012,nakajima2019imaging}, subsequently having an impact on the operation of graphene-based nanoscale electronic devices \cite{fernando2015oscillation, Katsnelson2008a, Zhang2020} as well as electrical characteristics of large-scale graphene samples. 

Out-of-plane disorder in graphene may occur for several reasons. For instance, graphene grown using the chemical vapour decomposition (CVD) process develops wrinkles and folds as a result of the thermal contraction of substrate during the cooling stage~\cite{DENG2016197, Wang2021,pan2011wrinkle}. The out-of-plane disorder may also be introduced during the transfer procedure~\cite{lanza2013tuning,liu2011origin}.  Significant efforts have then be devoted to eliminating wrinkles~\cite{dengWrinkleFreeSingleCrystalGraphene2017, Wang2021}, e.g. using the substrates with matching thermal expansion coefficients~\cite{lanza2013tuning}, strain engineering~\cite{hu2021eliminating} and tailored temperature control protocols~\cite{Wang2021}. Experimental studies of the electronic transport in graphene with out-of-plane disorder have also been published~\cite{Zhu2012,ma2020wrinkle}. It was proposed that controlled folding of graphene can be used for engineering charge-carrier dynamics~\cite{fan2021spontaneous,rode2018linking,folding_luo,Yang_origami}. No question, future applications of graphene in electronics call for a detailed understanding of the effect of this ubiquitous type of disorder on the electronic transport.

In this work, we systematically investigate the electronic transport across wrinkles and folds in graphene using first-principle computations. For commensurate graphene wrinkles, in which the interlayer stacking corresponds to the energetically favorable Bernal stacking configuration, we find that the electronic transmission oscillates over wide energy ranges. The observed oscillation patterns are attributed to quantum interference between the inter- and intralayer transport channels. In incommensurate wrinkles and folds, the mismatch between the layers is found to suppresses the interlayer tunneling resulting in transmission probabilities close to the limit of flat, pristine graphene. 
	
\section{Results}
	
\subsection{Construction of models}
	
The atomistic models of graphene with out-of-plane disorder considered in our work are defined by a compressive displacement of length $\Delta_W$ (see Fig.~\ref{fig:1}(a)) forming a wrinkle or a fold along  crystallographic vector $\vec{v}=(a,b)$. The considered configurations are thus assumed to be periodic along $\vec{v}$. The interplay between the bending energy and attractive interlayer interactions of graphene layers define the evolution across the three types of out-of-plane disorder realized upon increasing $\Delta_W$ as shown in Fig.~ \ref{fig:1}(b). While ripples are formed at small $\Delta_W$, interlayer attraction collapses such structures to wrinkles for larger values of $\Delta_W$, and further increase of $\Delta_W$ leads to folds, in which the contact area between graphene layers is further increased. Extremities of wrinkles and folds have loop-like structures free of interlayer coupling~\cite{Zhu2012}. All atomistic models of wrinkles and folds considered in our work have been constructed with the help of classical force-field relaxation (see the Methods section for details).
	
\subsection{Electronic transport across commensurate wrinkles}
	
We first consider the special case of wrinkles defined by $\vec{v}=(1,0)$ and $\vec{v}=(1,1)$, referring to them as zigzag and armchair, respectively. The collapsed regions of such wrinkles are compatible with the energetically favorable Bernal interlayer stacking configuration~\cite{lipson1942structure,butz2014dislocations,gargiulo2017structural,ni2014stacking}, and hence referred to as commensurate in the rest of our paper. For these relaxed models, we calculated ballistic charge-carrier transmission from first principles, using the combination of density functional theory (DFT) and the non-equilibrium Green's function formalism implemented in the TranSIESTA package~\cite{soler2002siesta,stokbro2003transiesta} (see Methods). The results of DFT calculations are discussed in comparison with the tight-binding (TB) approximation calculations employing the Slater-Koster formalism\ \cite{Zhu2012,SK_original} (see the Supplementary Materials). Figures~\ref{fig:armchair}(a)-(d) present the ballistic transmission $T(E,k_{//})$ for the models of zigzag wrinkles defined by $\Delta_W = 40,~60,~120$~and~240~\AA\  as a function of energy $E$ and momentum parallel to the wrinkle $k_{//}$.
Furthermore, each panel shows transmission $T(E)$ plotted at a specific $k_{//}$=$\pm 2\pi/(3a_0)$ ($a_0=2.46$~\AA\ is the lattice constant of graphene), which corresponds to the momentum of projections of the Dirac cone band degeneracies.
	
There are two striking observations in the presented transmission plots. Firstly, both in DFT and TB results, we observe a pronounced electron-hole asymmetry in the charge-carrier transmission.
The electron-hole asymmetry has an origin in the interlayer stacking of zigzag wrinkles. The collapsed region assumes Bernal stacking configurations AB$'_1$ or AB$'_2$ \cite{gilbert2019alternative}, as illustrated in Fig.~\ref{fig:1}(c), in which one of the graphene sublattices couples to itself upon folding since the two layers are mirror-symmetric with respect to each other. Such a coupling breaks the sublattice symmetry and hence the electron-hole symmetry~\cite{200662071,Semenoff_2012}.
	
	
	\begin{figure}
		\centering
		\includegraphics[width=1\linewidth]{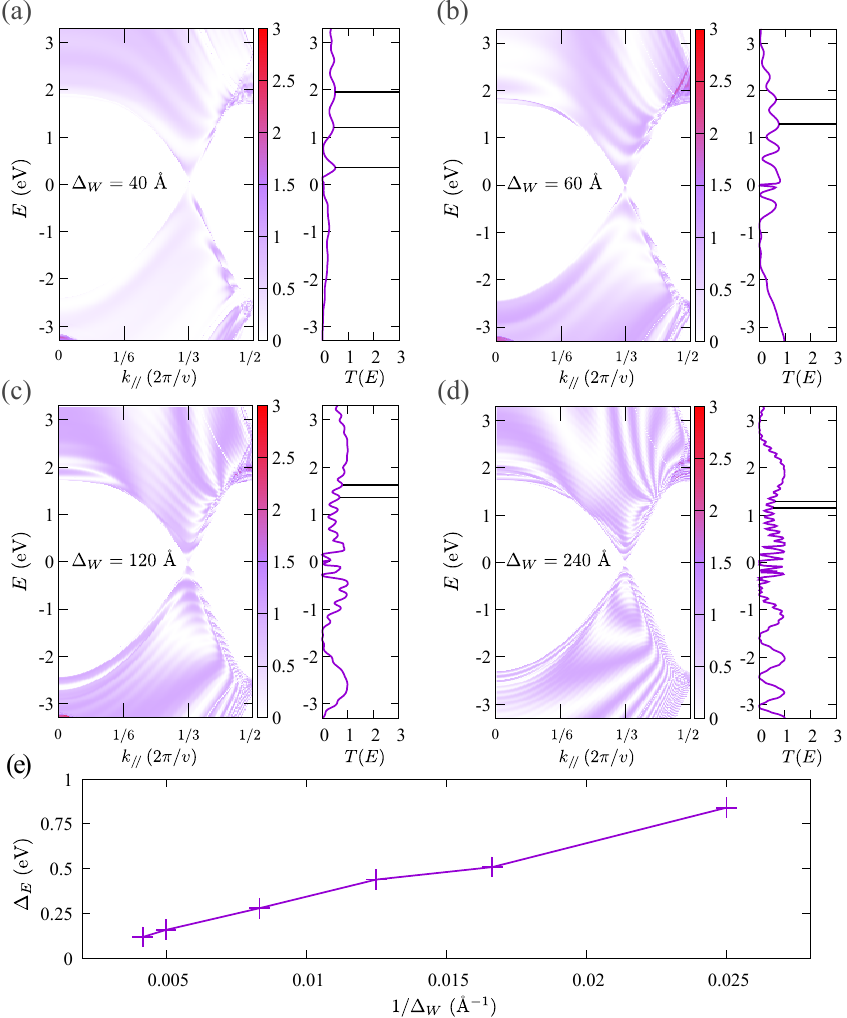}
		\caption{Ballistic transmissions $T(E,k_{//})$ of zigzag wrinkle models defined by (a)  $\Delta_W = 40$~\AA, (b) $\Delta_W= 60$~\AA, (c) $\Delta_W= 120$~\AA\ and (d) $\Delta_W=240$~\AA\ calculated from first principles. The side panels show the transmission probability at $k_{//} = 2\pi/(3a_0)$ which corresponds to the projections of the Dirac points.
        The energy spacing $\Delta_E$ between the oscillation peaks are highlighted by lines.
        (e) Dependence of $\Delta_E$ on $1/\Delta_W$.}
		\label{fig:armchair}
	\end{figure}
	
Secondly, ballistic transmission $T(E,k_{//})$ shows pronounced oscillations over broad energy ranges. Apart from making transmission highly energy-dependent, such oscillations also affect average conductance at a finite bias. These oscillations are clearly visible in the side panels of Figs.\ \ref{fig:armchair}(a-d) that show
transmission at a fixed momentum $k_{//} = 2\pi/(3a_0)$ that corresponds to the projections of the Dirac points.
Further analysis shows that the energy separation $\Delta_E$ between  the peaks has an approximately linear dependence on $\Delta_W$ (Fig.\ \ref{fig:armchair}(e)). Such a dependence is the signature of the interference between the interlayer and intralayer transport channels, as found by some of us previously in the case of electromechanical response of bilayer graphene \cite{fernando2015oscillation}.
This transport phenomenon is further addressed in Section\ \ref{chain}.

\begin{figure}
		\centering
		\includegraphics[width=1\linewidth]{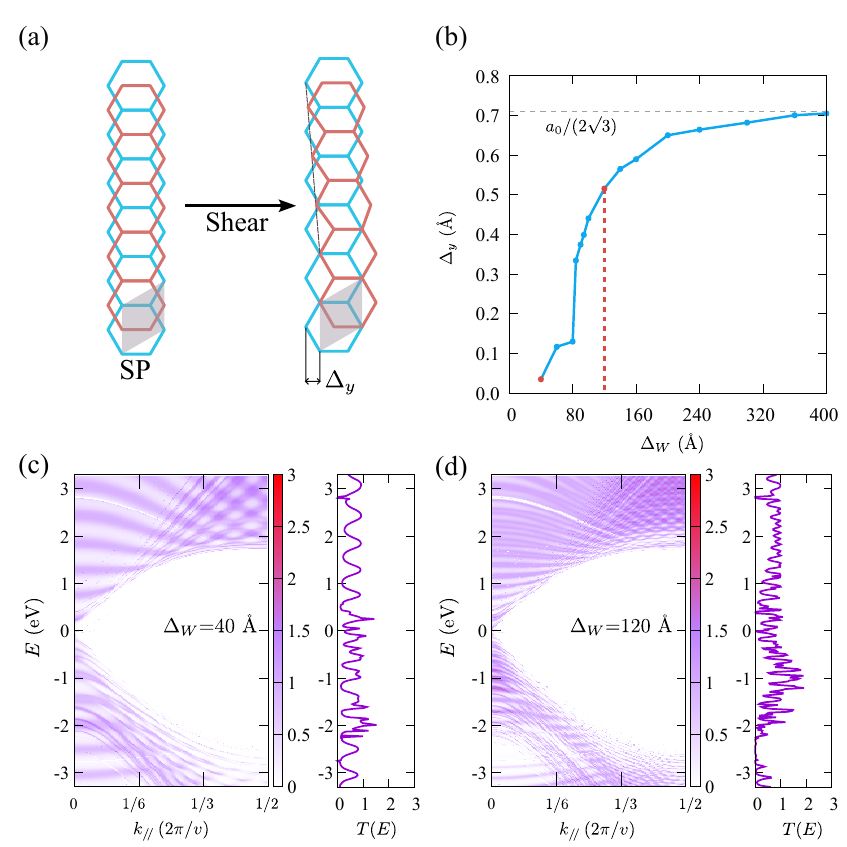}
		\caption{
		(a) Schematic illustration of the shear deformation in armchair wrinkles. The shear is characterized by displacement $\Delta_y$.
		(b) Evolution of shear deformation $\Delta_y$ versus compressive displacement $\Delta_W$. At small values of $\Delta_W$, shear deformation $\Delta_y$ is small, which corresponds to to the SP stacking configuration ($\Delta_y = a_0/(2\sqrt{3})$ corresponds to pure Bernal stacking.
		(c-d) Ballistic transmissions $T(E,k_{//})$ across armchair wrinkle models defined by (c) $\Delta_W=40$~\AA\ and (d) $\Delta_W=120$~\AA. The $T(E)$ cross sections are taken at $k_{//}$=0 that corresponding to the projected Dirac points. 
         }
		\label{fig:zigzag}
	\end{figure}

The second family of investigated commensurate configurations is defined by $\vec{v}=(1,1)$, that is wrinkles are oriented along the armchair direction. 
Atomic relaxation effects are more complex in such wrinkles.
Unlike in the zigzag case, realizing the lowest-energy Bernal stacking is possible only at a cost of introducing shear deformation as shown in Fig.~\ref{fig:zigzag}(a).
Consequently, the Bernal stacking is not achieved at small values of $\Delta_W$, and the collapsed region assumes the saddle-point (SP) stacking configuration~\cite{PSJ_stacking} that does not break sublattice symmetry.
Figure~\ref{fig:zigzag}(b) presents the evolution of shear deformation $\Delta_y$ upon the change of $\Delta_W$ with $\Delta_y = a_0/(2\sqrt{3})$ representing the pure Bernal stacking configuration.
Figures\ \ref{fig:zigzag}(c-d) present the transmission maps for the armchair wrinkles with $\Delta_W=40$~\AA\ and $\Delta_W=120$~\AA. 
In the case of $\vec{v}=(1,1)$, the Dirac points are projected onto  $k_{//}=0$. Similar to the case of zigzag wrinkles, oscillations with the $\Delta_E \propto 1/\Delta_W$ period are observed in the transmission maps. The oscillation pattern is more regular than in the case of $\Delta_W=40$~\AA\ armchair wrinkle, which assumes the SP stacking and hence preserves electron-hole symmetry. In contrast, the $\Delta_W=120$~\AA\ wrinkle is significantly closer to the Bernal stacking (see Fig.~\ref{fig:zigzag}(b)) and the electron-hole symmetry appears to be well visible in this case.

\subsection{Conductance oscillations in the atomic chain model}\label{chain}

	\begin{figure}
		\centering
		\includegraphics[width=1\linewidth]{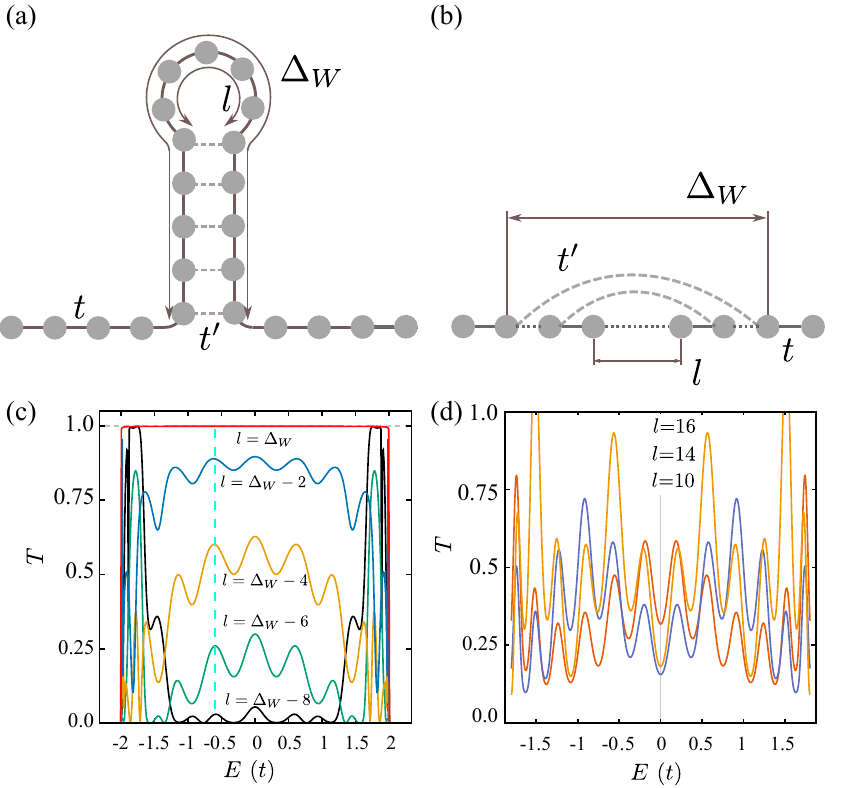}
		\caption{Transmission oscillations in atomic chain model. (a) Cross-section drawing of the trivialized graphene wrinkle and (b) its unfolded representation equivalent to atomic chain with additional hoppings. (c) Transmission $T$ as a function energy $E$ in units of $t$ calculated using the TB model Hamiltonian. In this plot $\Delta_W =12$ in units of intersite distance is fixed, while different curves correspond to difference values of $l$.  (d) First-order correction to the Green's function $\delta(E)=G_0\Delta h G_0/G_0 $ plotted for different $l$ and constant $\Delta_W=20$ reveals that the period of oscillations is governed by $\Delta_W$.}
		\label{fig:rainbow}
	\end{figure}

In order to further address the physical mechanism underlying the conductance oscillations observed in both the zigzag and armchair  wrinkles, we introduce a simple one-dimensional model treated using the tight-binding approximation. The presence of interlayer conductance channels is defined by $\Delta_W$, and also $l$ that represents the absence of interlayer hopping in the loop-like region as shown in Fig.~\ref{fig:rainbow}(a). At the same time, we observe that $k_{//}$ does not have any significant effect on the oscillation period, hence we introduce a one-dimensional chain described using the nearest-neighbor tight-binding model with an extra hopping $t'$  that models interlayer coupling in graphene wrinkles. Schematic diagram of this model with hopping $t'$ represented by a rainbow-like graph is shown in Fig.~\ref{fig:rainbow}(b). 
The ratio of the newly introduced hopping $t'$ to the nearest-neighbor hopping $t$ is chosen to resemble that of graphene wrinkles $t'/t = 0.48$~eV$/$$-$2.7~eV \cite{gargiulo2017structural,Zhu2012}. 
Figure\ \ref{fig:rainbow}(c) shows transmission $T$ as a function of energy $E$ at a fixed $\Delta_W$=12 in units of intersite distance, while parameter $l$ is varied. We observe that oscillation peaks have the same positions, which indicates that $l$ is of little effect on the oscillation period.
Combined with the results of DFT calculations we conclude that the oscillations are defined by the largest path difference $\Delta_W$.  
We further analyze the transmission oscillations in the atomic chain model using the non-equilibrium Green's functions (NEGF) approach, in which hoppings $t'$ are treated as a perturbative correction to the transmission.
	
First, we define an infinite atomic chain with the Hamiltonian
	\begin{align}
	H=t\sum_i c_{i}^{\dagger}c_{i+1}+h.c.,
	\label{chain_H}
	\end{align}
where $c_i$ $(c^{\dagger}_i)$ is the annihilation (creation) operator on the $i$th site. This Hamiltonian commutes with the translation operator, thus the energy eigenstates are also momentum eigenstates.
	
In the NEGF formalism~\cite{gargiulo2014topological,buttiker1986four}, the transmission is calculated as
	\begin{align}
	T(E)=\mathrm{Tr}[\Gamma_{1}G\Gamma_{2}G],
	\label{eq.trans}
	\end{align}
where $G$ is the Green's function $G(E)$=$[E-H-\Sigma]^{-1}$.
The coupling matrices $\Gamma_i$ are given by $\Gamma_i$=$i(\Sigma_i-\Sigma_i^{\dagger})$, with $\Sigma_i$ being the self-energies of the two semi-infinite leads.
	
Green's function $G_{0}$ describes the chain in absence of $t'$, while adding coupling $t'$ that models interlayer coupling in wrinkles adds an additional term $\Delta h$
	\begin{align}
	\Delta h=t' \sum_{i=l/2}^{\Delta_W/2}c_{i}^{\dagger}c_{-i}+h.c.
	\label{delta_h}
	\end{align}
The Green's function is then
	\begin{align}
	G(E)=&\frac{1}{G_{0}^{-1}-\Delta h} \nonumber \\
	=&G_{0}+G_{0}\Delta hG_{0}+G_{0}(\Delta hG_{0})^{2}+...~ .
	\label{Ge}
	\end{align}
Keeping only the first order of correction $G_{0}\Delta hG_{0}$, the transmission becomes
	\begin{align}
	T= & {\mathrm Tr}[\Gamma_{1}G_{0}\Gamma_{2}G_{0} \nonumber \\
	& +\Gamma_{1}G_{0}\Gamma_{2}(G_{0}+G_{0}\Delta hG_{0}) \nonumber \\
	&+\Gamma_{1}(G_{0}+G_{0}\Delta hG_{0})\Gamma_{2}G_{0} \nonumber \\
	& +\Gamma_{1}(G_{0}+G_{0}\Delta hG_{0})\Gamma_{2}(G_{0}+G_{0}\Delta hG_{0})].
	\label{T_pert}
	\end{align}
	
The Green's function can be written as an expansion involving eigenstates $|\psi_{n}\rangle$ of the chain with no hoppings $t'$ 
	\begin{align}
	G_{0}(E)=\sum_{n}\frac{1}{E+\varepsilon i-E_{n}}|\psi_{n}\rangle\langle\psi_{n}|,
	\label{g0_vec}
	\end{align}
and the correction term $G_{0}\Delta hG_{0}$ becomes
	\begin{align}
	G_{0}\Delta hG_{0}=\sum_{m}\sum_{n}\frac{|\psi_{m}\rangle\langle\psi_{m}|\Delta h|\psi_{n}\rangle\langle\psi_{n}|}{(E+\varepsilon i-E_{m})(E+\varepsilon i-E_{n})} .
	\label{g0_leading}
	\end{align}
As the simplest case, we analyze the $E_n=E_m$ correction $G_{0}\Delta hG_{0}=\langle\psi_{n}|\Delta h|\psi_{m} \rangle G_{0}$ that gives an $E_{i}$-dependent prefactor to the Green's function.
We write the factor as a function $\delta(E)$ as
	\begin{align}
	\delta(E)G_{0}=G_{0}\Delta hG_{0}.
	\end{align}
The leading order of transmission correction is $\Gamma_{1}(G_{0}+G_{0}\Delta hG_{0})\Gamma_{2}(G_{0}+G_{0}\Delta hG_{0})]$,
hence the correction to transmission contains $\delta^{2}+4\delta+1$.
	
We then evaluate the correction $\delta(E)$,\ keeping in mind that the eigenstates of the pristine chain
	\begin{align}
	\hat{H}|\psi(k)\rangle = 2t\cos(k)|\psi(k)\rangle,
	\label{correct_t}
	\end{align}
are also momentum eigenstates.
	The correction factor $\delta$ represents the phase difference between wavefunctions:
	\begin{align}
	    \delta(E)=\sum_{i}\langle\psi_{n}(r_{i})|\Delta_{h}|\psi_{n}(r_{-i})\rangle\big|_{E_{n}=E}
	\end{align} 
connected by the additional hoppings $t'$. It can then be approximated by a sum of sinusoidal
	functions
	\begin{align}
	\delta(k)=\frac{t'}{t}\sum_{i=l/2}^{\Delta_W/2}\cos(2ik).
	\label{fix}
	\end{align}
The results of the summation shown in Fig.\ \ref{fig:rainbow}(d) suggests that the highest-frequency component in Eq.~(\ref{fix}), which corresponds to the interference path $\Delta_W$, defines the oscillation peaks. Our first-principles results are consistent with the conclusions of this simple model. 
	
\subsection{Transport across incommensurate wrinkles}
	
We will now discuss graphene wrinkles formed along general crystallographic directions $\vec{v}=(a,b)$ other than high-symmetry zigzag and armchair orientations.
In these cases, the collapsed region locally forms twisted bilayer graphene with matching vectors $(a,b)$ and $(b, a)$.
The resulting twist angle is
	\begin{align}
	\theta = \arccos \left( \frac{a^2+4ab+b^2}{2(a^2+ab+b^2)} \right),
	\end{align}
while the translational vector along the wrinkle has a length of $d=\sqrt{a^2+b^2+ab}$.
	
We discuss the effect of wrinkle direction $(a,b)$ on the transmission $T(E,k_{//})$. Translational vector $(a,b)$ defines a one-dimensional mini Brillouin zone (mBZ) obtained by projecting the 2D Brillouin zone of graphene onto the $k_{//}$ direction in momentum space.
The Dirac cones of graphene are projected onto either $k_{//} =0$ (class Ia) or $k_{//} = 2\pi/(3|\vec v|)$ (class Ib) of the mBZ according to the classification introduced in  Ref.~\cite{yazyev2010electronic}. Class Ia is defined by $|a-b| \mod 3 = 0$, class Ib otherwise. The projections of the Dirac cones define the regions in the $T(E,k_{//})$ maps where transmission is allowed and limited by $n$ conductance channels in case of $n$-fold degeneracy of bands at given $E$ and $k_{//}$ in the ballistic regime.

\begin{figure}
		\centering
		\includegraphics[width=1.0\linewidth]{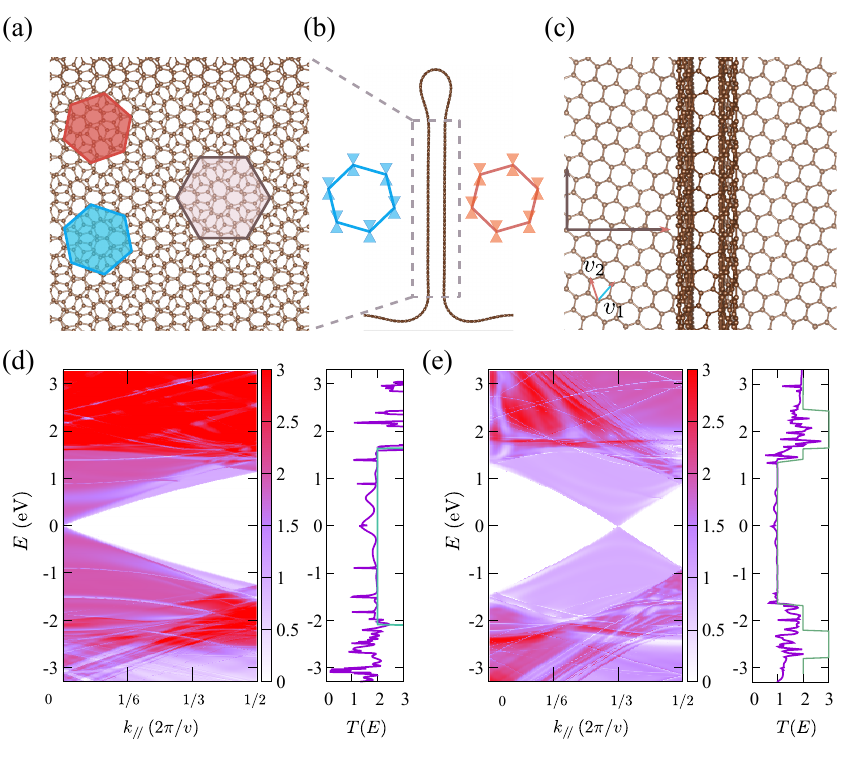}
		\caption{			
			Atomic structure of incommensurate wrinkle defined by the (1,2) direction: (a) local structure of the collapsed region equivalent to twisted bilayer graphene (unit cell is shown with the shaded region), (b) side-view with the sketch of the Brillouin zones and the Dirac cones of adjacent layers, and (c) top-view of the wrinkle illustrating the conservation of crystallographic orientation of graphene leads.
            Transmission maps $T(E,k_{//})$ for wrinkle models defined by (d) $\vec v$=(1,4) and $\Delta_W = 80$\AA (class Ia), (e)  $\vec v$=(1,2) and $\Delta_W = 80$\AA (class Ib).
		}
		\label{fig:twist_wrinkle}
	\end{figure}

The periodic structure of wrinkles results in consequences deeper than just the conservation of momentum $k_{//}$ upon ballistic transmission. We stress that semi-infinite graphene sheets on both sides of wrinkles of constant width have the same crystallographic orientation. The momentum conservation implies suppressed backscattering at the Dirac point, which can be observed by evaluating contribution to the transmission from the first-order correction $G_0 \Delta h G_0$.
Starting with the pristine graphene and a simple interlayer containing only hopping between aligned atoms
	\begin{align}
	\Delta h_{ij}=\begin{cases}
	t', & r_{i}^{\perp}=r_{j}^{\perp}\\
	0, & r_{i}^{\perp}\neq r_{j}^{\perp},
	\end{cases}
	\end{align}
the effective $\Delta G$ writes
	\begin{align}
	G_0 \Delta h G_0(z) = \sum_m\sum_n\frac{\langle\psi_{m}|\Delta h|\psi_{n}\rangle}{(z-E_m)(z-E_n)}|\psi_{m}\rangle\langle\psi_{n}|,
	\end{align}
which becomes most significant at $E_m=E_n=z$.
Recalling the fact that $|\psi_m\rangle$ and $|\psi_n\rangle$ are eigenstates of pristine graphene, 
	$\langle\psi_{m}|\Delta h|\psi_{n}\rangle$ gives an $\exp(2\pi i (\vec{k}_m-\vec{k}_n)\cdot\vec{r}_{ij})$ term.
Integrating over $\vec{r}_{ij}$, $\Delta G$ vanishes if $\vec{k}_m \neq \vec{k}_n$, while the wrinkle enforces a transformation $\vec{k}_m = \mathcal{M}_x \vec{k}_n$ due to its mirror-symmetric stacking configuration of the two layers as shown in Figs.\ \ref{fig:twist_wrinkle}(a,b). Here, $\mathcal{M}_x$ denotes the mirror-reflection with respect to transport direction $x$: $\mathcal{M}_x (k_x,k_y) = (-k_x,k_y)$.
From the above rules of momentum conservation, 
we conclude that the transmission is only affected in the overlapping region of the Dirac cones.
In the non-overlapping region, the correction $G_0 \Delta h G_0$ is vanishing, and the transmission retains the value of ideal, defect-free graphene.
These results are verified by the explicit DFT transport calculations as shown in Fig.~\ref{fig:twist_wrinkle}(d-e) for class Ia and class Ib wrinkles, respectively. 
The transmission maps $T(E,k_{//})$ have overall shape of the Dirac cone projections. Transmission values near the charge neutrality are $T\approx 2$ and $T \approx 1$ for class Ia and Ib configurations, respectively, indicating that interlayer tunnelling plays a minor role. At higher energies where the Dirac cones overlap, e.g. near $E\approx2$\ eV in Fig.\ \ref{fig:twist_wrinkle}(e), backscattering becomes significant leading to a series of transmission dips.
We also point out that class Ia presents larger backscattering from the interlayer coupling since the projected Dirac cones overlap with each other.

\subsection{Transport across graphene folds}
	
\begin{figure}
		\centering
		\includegraphics[width=\linewidth]{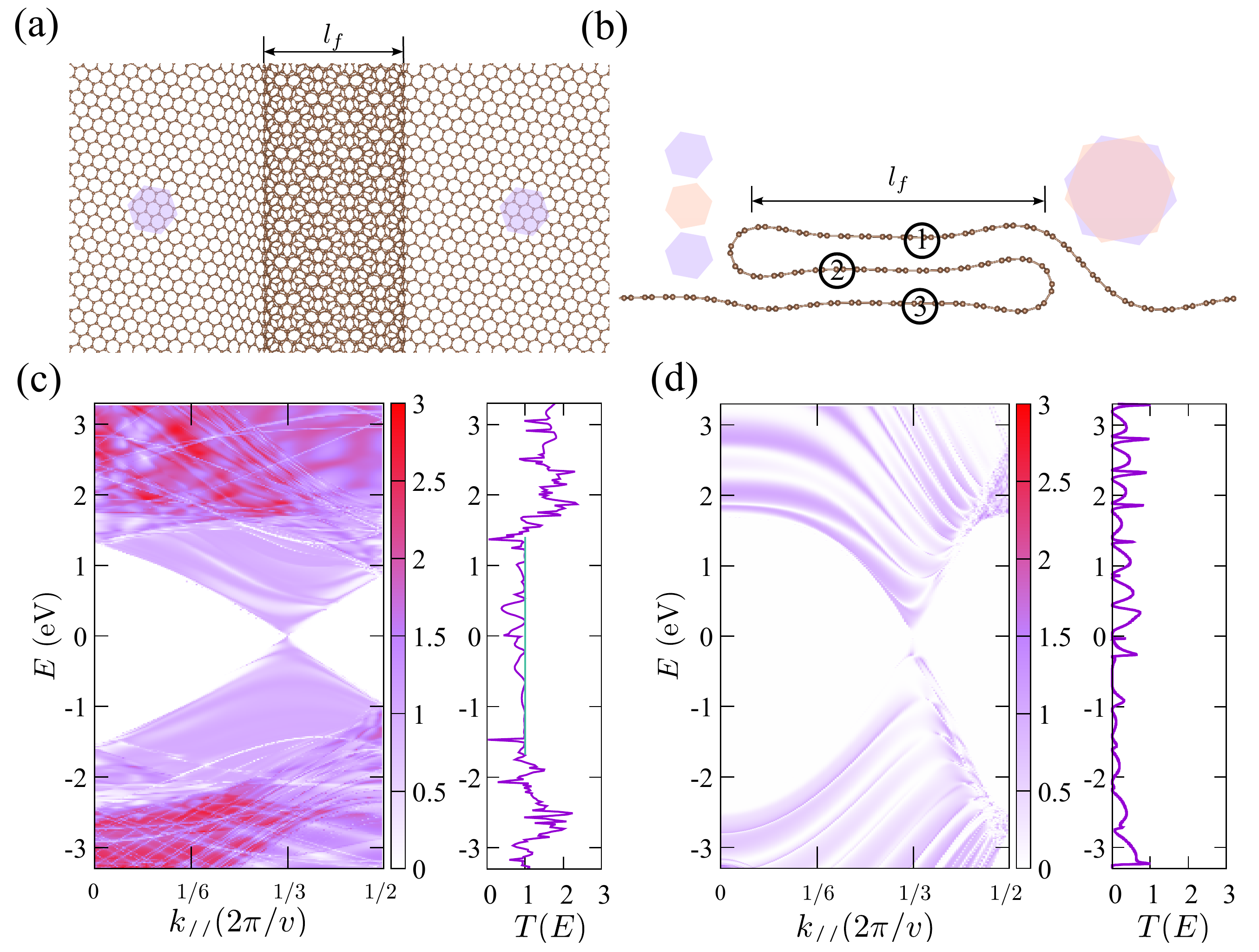}
		\caption{(a) Atomic structure of an incommensurate fold defined by  $\vec{v}=$~(1,2) as an example. 
		(b) Side-view of the fold with layers numbered and Brillouin zone orientations indicated. Transmission maps $T(E,k_{//})$ for (c) the incommensurate fold shown in the above panels and (d) zigzag fold characterized by $\Delta_W = 80\ $\AA
		.}
		\label{fig:transkfold}
	\end{figure}

We will now discuss folds as the ultimate regime of out-of-plane disorder in graphene. Folds realize triple-layer graphene configurations in their collapsed regions (Fig. \ref{fig:transkfold}(a-c)).  Importantly, adjacent layers (pairs 1--2 and 2--3) in incommensurate folds are twisted with respect to each other, while the outside layers 1 and 3 are aligned. This configuration is equivalent to mirror-symmetric twisted trilayer graphene.
While we still expect the effect of interlayer coupling to be weakened by the incommensuration, our DFT calculations predict a larger degree of backscattering in folds than in wrinkles (compare Figs.~\ref{fig:twist_wrinkle}(e) and \ref{fig:transkfold}(d) for the the (1,2) direction).
For the folded region of width $l_f=40$~\AA, the average transmission in the energy interval ($-$0.15\ eV,~0.15\ eV) is 0.727, while in the wrinkle of equivalent $\Delta_W = 80$~\AA\ it is 0.908. 
The observed transport behaviour raises the question of whether the enhanced backscattering in incommensurate folds as compared to wrinkles originates from the direct coupling of the outmost layers 1 and 3.
The corresponding matrix elements of the Hamiltonian in localized-basis-set first-principles calculations \cite{soler2002siesta,zerothi_sisl}, 
are found to be negligible. The estimated Slater-Koster coupling also has a negligible magnitude of $10^{-4}$~eV.
Therefore, we attribute the enhanced scattering to the fact that the number of interlayer tunneling channels is doubled in the folds.
As expected, for a commensurate zigzag fold (Fig. \ref{fig:transkfold}(d)) we observe strong backscattering with transmission magnitudes lower than in the equivalent zigzag wrinkles (Fig.~\ref{fig:armchair}).
	
	\section{Discussion}
We investigated the effect of our-of-plane disorder on the electronic transmission in graphene.
Different forms of the  our-of-plane disorder exist in graphene, depending on the compressive displacement and the orientation of the deformation. 
Our work studied ballistic transmission through the wrinkles and folds using first-principle calculations, taking into account their width and interlayer commensuration.
	
The interlayer coupling was found to cause substantial oscillations in the electronic transmission across commensurate wrinkles.
Such oscillations were found to originate from the quantum interference involving the interlayer tunneling channels.
Based on DFT calculations, we propose a simple one-dimensional model that fully captures the observed oscillations.
On the other hand, in  incommensurate, ``twisted'' wrinkles the interlayer coupling is effectively weaker, and the transmission near the Fermi level preserves that of pristine, flat graphene.
We have also found enhanced backscattering in folds that was attributed to the doubled contact region in this type of the out-of-plane disorder.

Our results offer an approach toward understanding the transport in mesoscopic graphene samples containing out-of-plane disorder of different type and arbitrary orientation.
The theory of transmission across graphene wrinkles and folds is thus useful for designing graphene-based devices as well as fold-engineering of graphene.
As a generalization, the principles presented in our work are expected to apply also to other types of 2D materials.
Formation of locally twisted bilayers in the wrinkles and folds provides an interesting outlook for further studies, e.g. the ``twisted'' wrinkles in the smaller-angle regime.
	
\section*{Acknowledgements}
This work was supported by the Swiss National Science Foundation (grant No. 172543). Computations were performed at the Swiss National Supercomputing Centre (CSCS) under project No. s1146 and the facilities of Scientific IT and Application Support Center of EPFL.
	
\section*{Methods}
	
\subsection{Structure relaxation with classical force fields}
	
The atomic structures of models of the out-of-plane disorder in graphene were obtained by means of the classical force field simulations using LAMMPS\ \cite{LAMMPS,LammpsJCP}. The classical force field includes the bond-order potential for describing covalent bonding~\cite{SRpotentional} as well as the modified version of the Kolmogorov–Crespi registry-dependent potential~\cite{KolmogorovPRB} for describing the interlayer van der Waals  interactions.
The energy minimization was performed using the conjugate-gradient and fire algorithms.
	
\subsection{The tight-binding model calculations}
\label{tb_methods}
	
In order to describe both the interlayer coupling and the effect of curvature in the tight-binding calculations of graphene with out-of-plane disorder, we employ the Slater-Koster model~\cite{SK_original,Zhu2012}.
The $p_z$ atomic orbitals of carbon atoms form the intralayer $\pi$ bonds and the interlayer $\sigma$ bonds.
The general form of the Hamiltonian including both contributions is
	\[
	\hat{H} = \sum_{i,j}^{} t^{ij}_{\pi} c^{\dagger}_i c_j + \sum_{i,j}^{} t^{ij}_{\sigma} c^{\dagger}_i c_j.
	\]
Explicit expressions for the hoppings $t^{ij}_{\pi}$ and $t^{ij}_{\sigma}$ are \ \cite{Zhu2012}
	\begin{align}
	t^{ij}_{\pi}=V_{\pi}^{0}\exp(-\frac{r-a_{0}}{r_{0}})|\sin\theta_{i}\sin\theta_{j}|, \\
	t^{ij}_{\sigma}=V_{\sigma}^{0}\exp(-\frac{r-d_{0}}{r_{0}})|\cos\theta_{i}\cos\theta_{j}|.
	\end{align}
Following the previous\ \cite{Zhu2012} Slater-Koster parametrization, we set $V_{\pi}^0=-2.7$~eV, $V_{\sigma}^{0}=0.48$~eV, characteristic distances $a_0=1.42~\mathrm{\AA}$, $d_0=3.35~\mathrm{\AA}$ and the decay length $r_0=0.184a$ ($a=\sqrt{3}a_0=2.46~\mathrm{\AA}$  as defined in the main text).
In the orientation-dependent terms, angles $\theta_i$ and $\theta_j$ are defined as the angle between $\vec{r}_{ij}$ and the local normal vector at atomic positions $\vec{r}_{i}$ and $\vec{r}_{j}$, that is $\theta_i = \angle (\vec r_{ij}, \vec n_i)$.
These terms accounts for the effect that the local curvature of graphene sheet on the overlap between $p_z$ orbitals.
	
\subsection{Recursive Green's function methods}
\label{NeGF}

The ballistic transmission was calculated using the non-equilibrium Green's function methods in both the TB model and DFT calculations.
The transmission probability is expressed as
	\begin{align}
	T_{ij}(E)=\mathrm{Tr} [\Gamma_i G \Gamma_j G],
	\end{align}
where $G$ is the Green's function as used in Eq.~(\ref{eq.trans}): $G=[G_0-\Sigma]^{-1}$.
The $\Gamma$ matrices contains the self-energy terms of the two leads 
	\begin{align}
	\Gamma_i(E) = i[\Sigma_i(E) -\Sigma_i^{\dagger}(E)] .
	\end{align}
The self-energy from the $i$th lead is calculated as $\Sigma_i = h_i G_i h_i^\dagger$, where $h_i$ is the coupling matrix between the lead and the scattering region. For each of the semi-infinite leads, Green's function $G_i$ is obtained through the recursive Green's function methods. In each step one layer is added to the lead, and the Green's function iterates as $g_j$=$[E-h-Tg_{j-1}T^\dagger]^{-1}$. 
$G_i$ is taken as the converged value of $g$, that is $G_i = g_j^{j\rightarrow\infty}$.
	
\subsection{First-principles electronic transport calculations}

First-principles transport calculations were performed with TranSIESTA package\  \cite{soler2002siesta,stokbro2003transiesta}.
We used the double-$\zeta$ plus polarization basis set combined with the local density approximation exchange-correlation functional~\cite{LDA_PRB}.
The energy shift for constructing the localized basis was set to 275~meV, and the real-space cutoff to 250~Ry.
The estimation of the direct coupling between the top and bottom layers in graphene folds was extracted from the localized basis set Hamiltonian using the sisl package~\cite{zerothi_sisl}.

	\bibliography{fold}

\end{document}